\documentclass[fleqn,10pt]{wlscirep}
\usepackage[utf8]{inputenc}
\usepackage[T1]{fontenc}

\usepackage{amsmath}
\usepackage{graphicx}
\usepackage{dcolumn}
\usepackage[english]{babel}

\usepackage{color}
\usepackage{float}

\title{Effect of carbon content on electronic structure of uranium carbides}

\author[1,*]{Sergei M. Butorin}
\author[2,3]{Stephen Bauters}
\author[2,3]{Lucia Amidani}
\author[4]{Aaron Beck}
\author[2,3]{Andr\'{e} Rossberg}
\author[2]{Stephan Weiss}
\author[4]{Tonya Vitova}
\author[2,3]{Kristina O. Kvashnina}
\author[5]{Olivier Tougait}

\affil[1]{Condensed Matter Physics of Energy Materials, X-ray Photon Science, Department of Physics and Astronomy, Uppsala University, P.O. Box 516, SE-751 20 Uppsala, Sweden}
\affil[2]{Helmholtz-Zentrum Dresden-Rossendorf (HZDR), Institute of Resource Ecology, P.O. Box 510119, 01314, Dresden, Germany}
\affil[3]{The Rossendorf Beamline at ESRF-The European Synchrotron, 38043 Grenoble, France}
\affil[4]{Institute for Nuclear Waste Disposal (INE), Karlsruhe Institute of Technology, P.O. 3640, D-76021 Karlsruhe, Germany}
\affil[5]{Univ. Lille, CNRS, Centrale Lille, Univ. Artois, UMR 8181 - UCCS - Unit\'{e} de Catalyse et Chimie du Solide, F-59000 Lille, France}
\affil[*]{sergei.butorin@physics.uu.se}

\begin{abstract}
The electronic structure of UC$_x$ (x=0.9, 1.0, 1.1, 2.0) was studied by means of x-ray absorption spectroscopy (XAS) at the C $K$ edge and measurements in the high energy resolution fluorescence detection (HERFD) mode at the U $M_4$ and $L_3$ edges. The full-relativistic density functional theory calculations taking into account the $5f-5f$ Coulomb interaction $U$ and spin-orbit coupling (DFT+$U$+SOC) were also performed for UC and UC$_2$. While the U $L_3$ HERFD-XAS spectra of the studied samples reveal little difference, the U $M_4$ HERFD-XAS spectra show certain sensitivity to the varying carbon content in uranium carbides. The observed gradual changes in the U $M_4$ HERFD spectra suggest an increase in the C $2p$-U $5f$ charge transfer, which is supported by the orbital population analysis in the DFT+$U$+SOC calculations, indicating an increase in the U $5f$ occupancy in UC$_2$ as compared to that in UC. On the other hand, the density of states at the Fermi level were found to be significantly lower in UC$_2$, thus affecting the thermodynamic properties. Both the x-ray spectroscopic data (in particular, the C $K$ XAS measurements) and results of the DFT+$U$+SOC indicate the importance of taking into account $U$ and SOC for the description of the electronic structure of actinide carbides.
\end{abstract}

\begin{document}

\flushbottom
\maketitle
% * <john.hammersley@gmail.com> 2015-02-09T12:07:31.197Z:
%
%  Click the title above to edit the author information and abstract
%
\thispagestyle{empty}

%/////////////////////////////////////////////////////////
%------------------- Introduction ------------------------
%/////////////////////////////////////////////////////////
\section*{Introduction}
Actinide carbides are considered as advanced nuclear fuels for the Generation IV nuclear reactors which will allow for the transmutation of minor actinides, thus contributing to the challenge of utilizing the nuclear waste. In turn, the performance of the carbide fuel will depend on the stoichiometry and stability of the required phases of the carbide systems. The thermodynamic properties will also depend on the changes in the electronic structure of actinide carbides. Since the main material for the fuel is uranium carbide, the U-C system receives more attention in terms of the dedicated research as compared to other actinide carbides.

X-ray spectroscopy is a good tool to probe the electronic structure of different materials but for uranium carbides most of the studies were so far carried out with the help of conventional x-ray spectroscopic techniques, such as x-ray absorption spectroscopy (XAS) at the U $L_3$ edge \cite{Itie,Vigier,Carvajal} or x-ray photoemission spectroscopy (XPS) \cite{Erbudak,Ishii,Ito,Ejima,Eckle,Dillard,Ejima02,Jilek}. A great improvement in the quality of such research of actinide systems came with the application of the high energy resolution fluorescence detection x-ray absorption spectroscopy (HERFD-XAS) at the actinide $M_{4,5}$ edges \cite{Kvashnina01,Butorin01,Vitova01} (see also reviews \cite{Kvashnina02,Kvashnina03}) when the greatly enhanced energy resolution and sensitivity of the method helped to resolve some long standing questions \cite{Kvashnina01} about the oxidation path, monitor a gradual change of the oxidation state upon doping \cite{Butorin02}, discover new phases \cite{Kvashnina04} and clarify the mechanism of the nano-phase formation \cite{Pan}. The followed theoretical efforts in modelling the HERFD-XAS spectra for the large part of the actinide row \cite{Butorin03,Butorin04} made the interpretation the experimental data much easier.

Earlier, we reported the results of the HERFD-XAS measurements on UC and UMeC$_{2}$ (Me=Fe, Zr, Mo) carbides \cite{Butorin05}. Here we report the data measured for the uranium carbide samples with different carbon content: UC$_{0.9}$, UC, UC$_{1.1}$ and UC$_{2}$.

%/////////////////////////////////////////////////////////
%-------------- Results and discussion -------------------
%/////////////////////////////////////////////////////////
\section*{Results and discussion}
\begin{figure}
\includegraphics[width=\columnwidth]{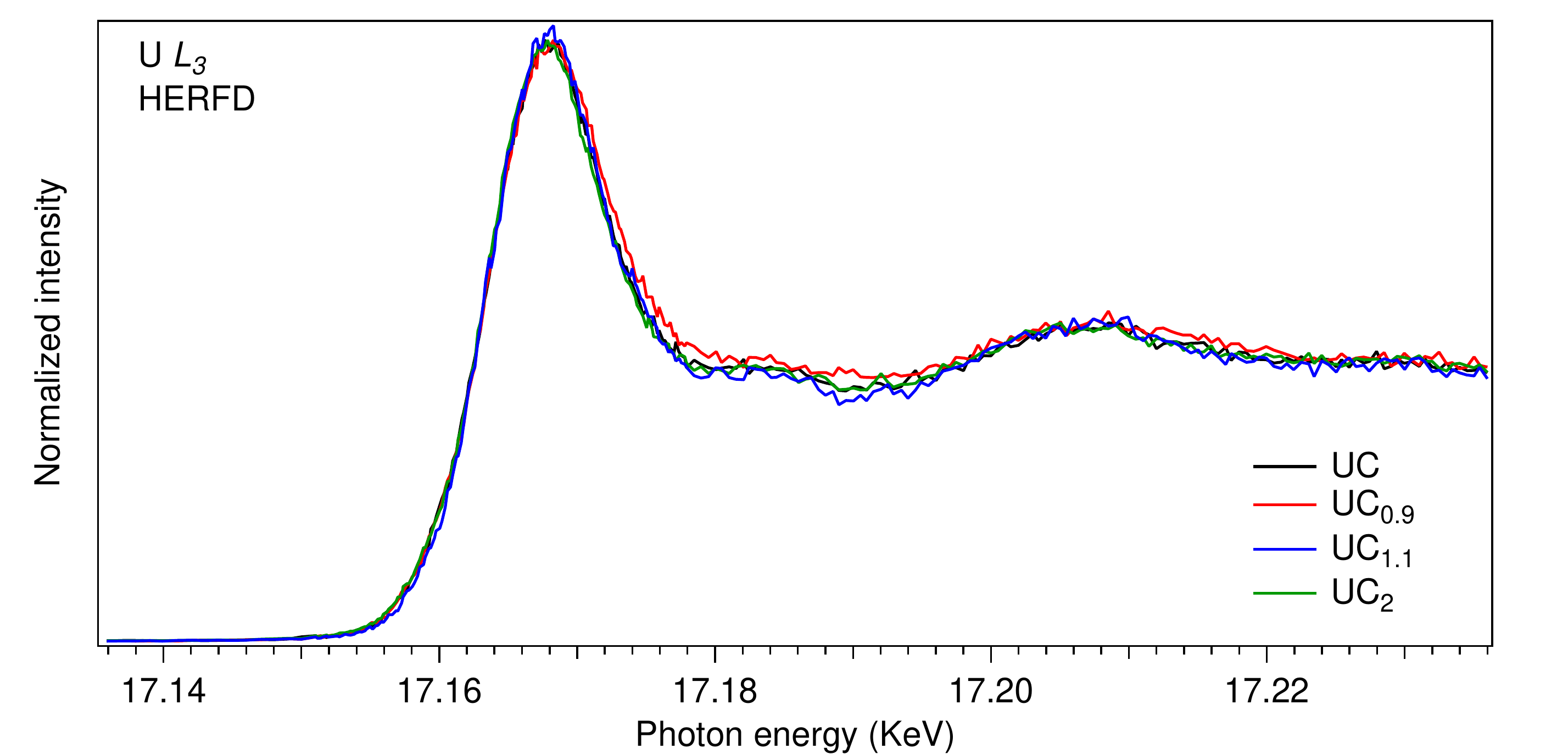}
\caption{U $L_3$ HERFD-XAS of UC$_{0.9}$, UC, UC$_{1.1}$ and UC$_{2}$.}
\label{UL3_HERFD}
\end{figure}

Fig.~\ref{UL3_HERFD} displays the HERFD-XAS spectra of the UC$_{0.9}$, UC, UC$_{1.1}$ and UC$_{2}$ samples recorded at the U $L_3$ edge. The spectra do not reveal significant changes between the samples. For UC$_{0.9}$, a slight increase of the spectral intensity is observed in the 17172-17195 eV energy region as compared to the spectra of other carbide compositions which is consistent with x-ray diffraction (XRD) data indicating the existing mixture of UC and $\alpha$-U in the UC$_{0.9}$ sample. The post-white-line background is expected to be higher for pure uranium than for its carbide \cite{Vigier}.

\begin{figure}
\includegraphics[width=\columnwidth]{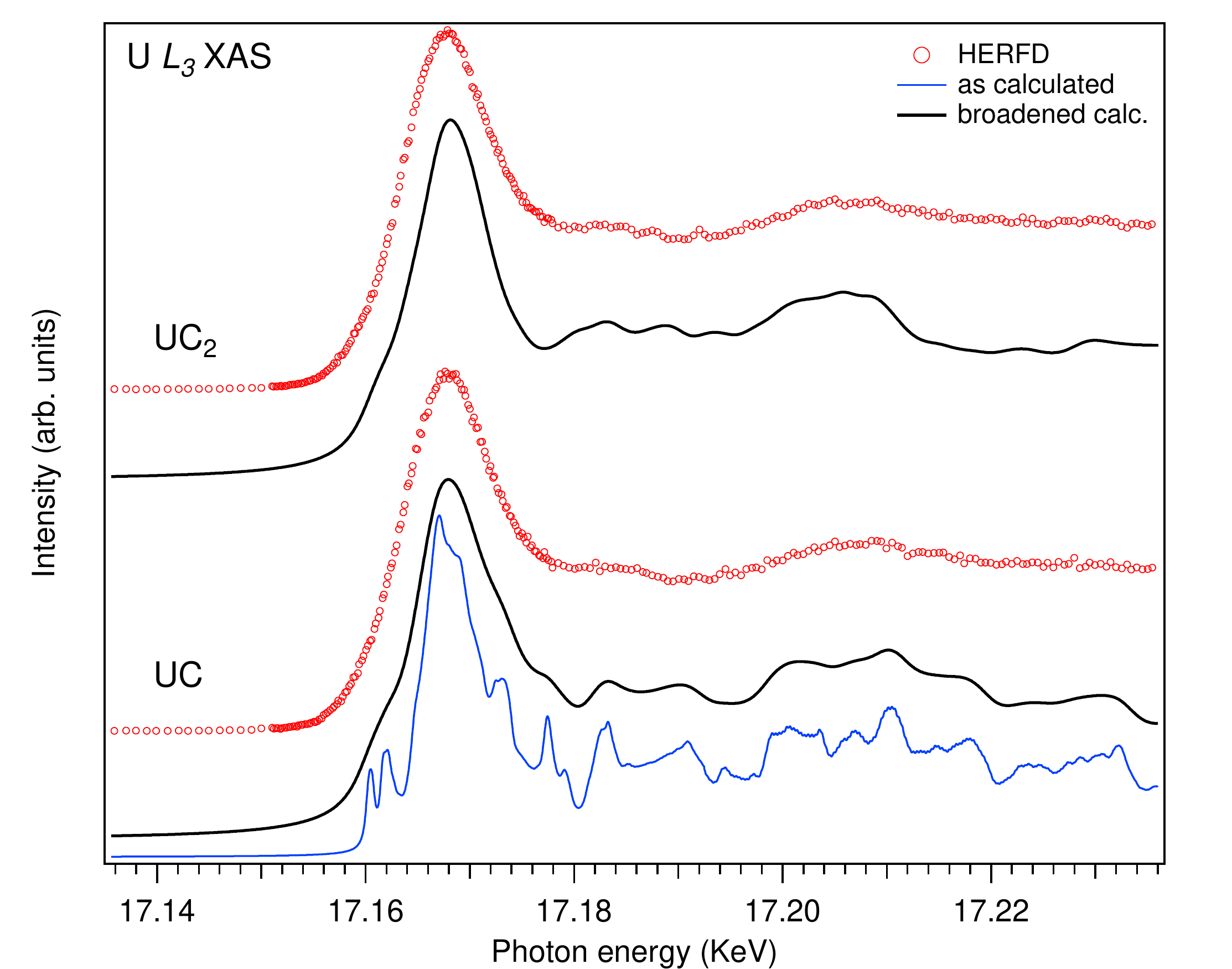}
\caption{Experimental U $L_3$ HERFD-XAS spectra of UC and UC$_2$ compared to BSE-calculated ones using OCEAN code with DFT+$U$ approach.}
\label{UL3_compare}
\end{figure}

To evaluate the underlying spectral structures, Fig.~\ref{UL3_compare} compares the experimental U $L_3$ HERFD-XAS spectra of UC and UC$_2$ with the results of full-relativistic density functional theory taking into account the $5f-5f$ Coulomb interaction $U$ (DFT+$U$) and Bethe-Salpeter equation (BSE) based calculations of U $L_3$ XAS using the OCEAN code and crystal structures from Ref. \cite{Austin}. For the illustration purpose, both small and large broadenings of the calculated U $L_3$ XAS spectrum of UC were used to show the contribution of various underlying structures. Besides the dipole $2p\rightarrow{d,s}$ transitions (the $s$ contribution is small), the quadrupole $2p\rightarrow{f}$ transitions were also calculated. The latter give the main contribution to the low energy side of the U $L_3$ XAS spectra (see structures around 17162 eV in Fig.~\ref{UL3_compare}). These quadrupole transitions are significantly affected by taking into account $U$ for the $5f$ shell. The dipole transitions to the $d$ states are also affected to some extent due to a modified $6d$ density of states (DOS) as a result of the $6d$-$5f$ hybridization. The $U$ value for UC was discussed in several publications \cite{Shi,Ducher,Wdowik} and most of the ground state and spectroscopic properties were reproduced using the DFT+$U$ approach with $U$=2.0-3.0 eV. Therefore, both for UC and UC$_2$, we used the $U$=2.5 eV value in the calculations and $J$ was set to 0.5 eV. While the post-edge structures of the $L_3$ spectra are usually sensitive to changes in the crystal structure \cite{Vitova02,Kvashnina02}, in this case, even an improved energy resolution of the HERFD mode does not allow to make a clear distinction between cubic UC and tetragonal UC$_{2}$.

A significant broadening of the U $L_3$ XAS (and even HERFD) spectra, which mainly probe the unoccupied $6d$ states, makes it difficult to study the changes in the U $6d$ DOS in detail. XAS measurements at the U $N_{6,7}$ edges ($4f\rightarrow6d$ transitions) can provide a higher energy resolution \cite{Butorin06} but are not easy to perform due to weak intensities. XAS measurements at the C $K$ edge, which probe the unoccupied C $2p$ states, can also help with this due to an admixture of the C $2p$ states to the $5f$ and $6d$ states as a result of the U $5f$-O $2p$ and U $6d$-O $2p$ hybridization. As it was shown earlier, the ligand/anion $K$ XAS spectra of actinide materials are sensitive to changes in the crystal structure of materials as well as various non-stoichiometry and defects \cite{Modin}. Therefore, the detailed understanding of the nature of the features in the ligand $K$ XAS spectra is necessary in order to use them for the material characterization.

\begin{figure}
\includegraphics[width=\columnwidth]{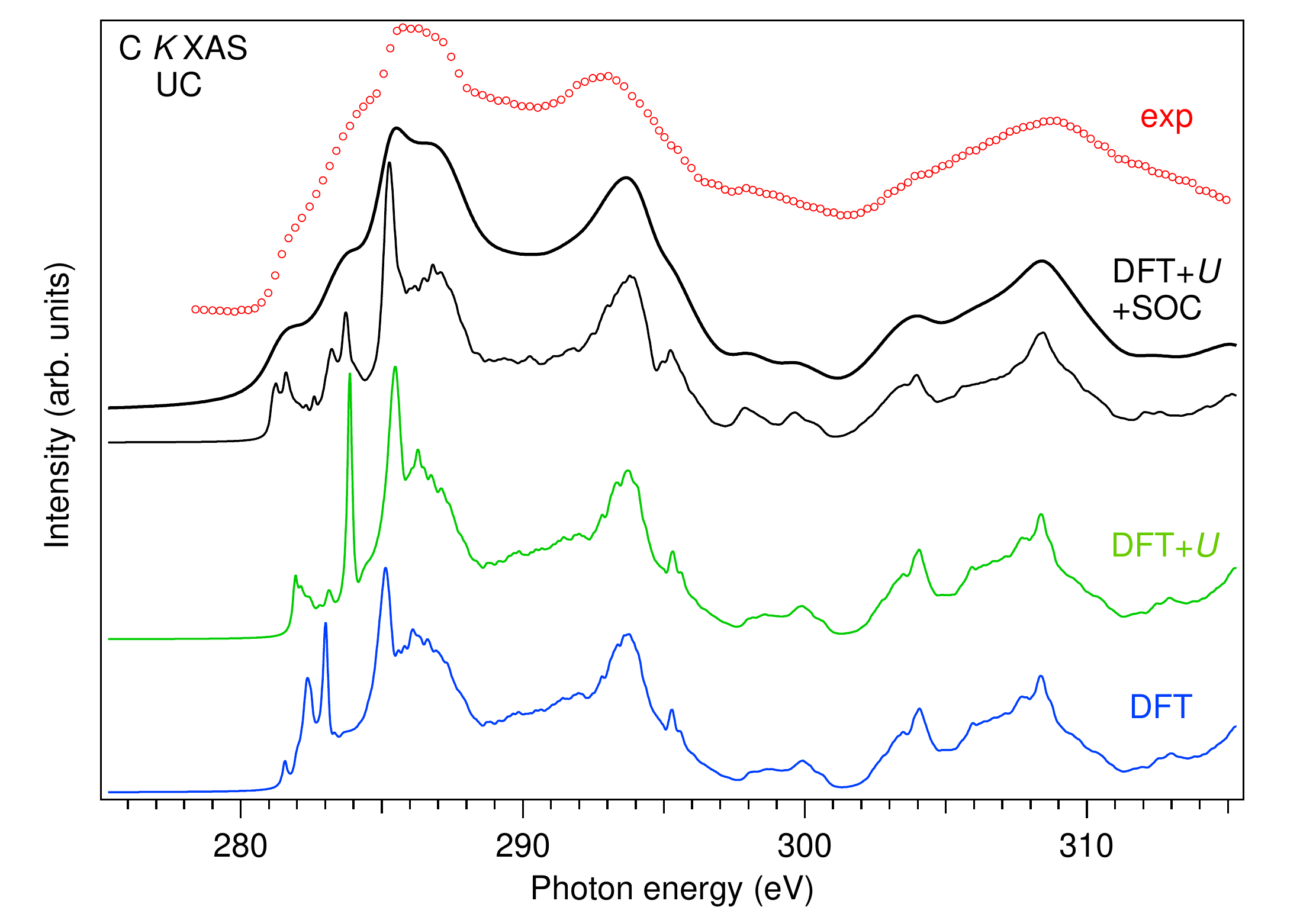}
\caption{For UC, experimental C $K$ XAS spectrum (red markers) compared to BSE-calculated ones using OCEAN code with DFT (blue curve), DFT+$U$ (green curve) and DFT+$U$+SOC (black curve) approaches, respectively.}
\label{CK_UC}
\end{figure}

Fig.~\ref{CK_UC} displays the experimental C $K$ XAS spectrum of UC which is compared to BSE-calculated C $K$ XAS spectra using various formalisms: DFT, DFT+$U$ and DFT+$U$ with taking into account spin-orbit coupling (SOC). The same $U$ and $J$ values were used as in the case of the U $L_3$ XAS calculations. In the DFT+$U$+SOC case, the optimal projector functions were calculated with scalar-relativistic versions of U and C pseudopotentials, while full-relativistic pseudopotentials were used for other stages of the calculations in the OCEAN code. The broadening of the calculated spectra is limited to only the core-hole lifetime to get better understanding of the underlying structures in the experimental spectrum. In the DFT+$U$+SOC case, the calculated spectrum was additionally broadened to take into account the instrumental resolution for a better comparison with experiment.

In the DFT-based calculations (blue curve in Fig.~\ref{CK_UC}), the low energy structure at $\sim$282.6 eV corresponds to transitions to the C $2p$ states hybridized to the U $5f$ states while the structure at $\sim$286.1 eV is associated with C $2p$ states admixed to the U $6d$ states. An inspection of Fig.~\ref{CK_UC} reveals a significant influence of the $5f-5f$ Coulomb interaction $U$ and SOC on the shape of the C $K$ XAS spectrum of UC. While the $5f-5f$ Coulomb interaction only affects the C $2p$ states hybridized with the U $5f$ states (see green curve in Fig.~\ref{CK_UC}), SOC also affects the C $2p$ states hybridized with the U $6d$ states (black curves in Fig.~\ref{CK_UC}). Compared to the scalar-relativistic DFT calculations (blue curve), SOC leads to a significant splitting of the $\sim$282.6-eV structure into two groups at $\sim$281.5 and $\sim$283.4 eV in the calculated C $K$ XAS spectrum, while a change of the $\sim$286.1-eV structure, reflecting a modification of the U $6d$ states, is less pronounced. Therefore, the low energy structures of the experimental C $K$ XAS spectrum can be considered as a C $2p$ manifestation of $U$ and SOC acting on the U $5f$ states, thus indicating the importance of these interactions for the electronic structure characterization of actinide carbides.

In contrast to the U $L_3$ HERFD-XAS data, the U $M_4$ HERFD-XAS spectra of uranium carbides, which probe the U $5f$ states, are turned to be sensitive to the carbon content and varying composition of the samples. In particular, the shoulder on the high energy side of the U $M_4$ HERFD spectrum at $\sim$3727 eV grows with increasing C content in the composition. It is difficult to explain the shoulder growth by only a change in the multiplet structure of the ground state $5f^{3}$ configuration \cite{Butorin05} as a result of the crystal structure distortion. If UC$_{0.9}$ is a mixture of $\alpha$-U and UC and UC$_{1.1}$ is a mixture of UC and UC$_2$, the gradual relative-intensity increase of the shoulder upon going from UC$_{0.9}$ to UC then to UC$_{1.1}$ and to UC$_2$ can not be caused by the low symmetry $\rightarrow$ high symmetry $\rightarrow$ low symmetry transition.

\begin{figure}
\includegraphics[width=\columnwidth]{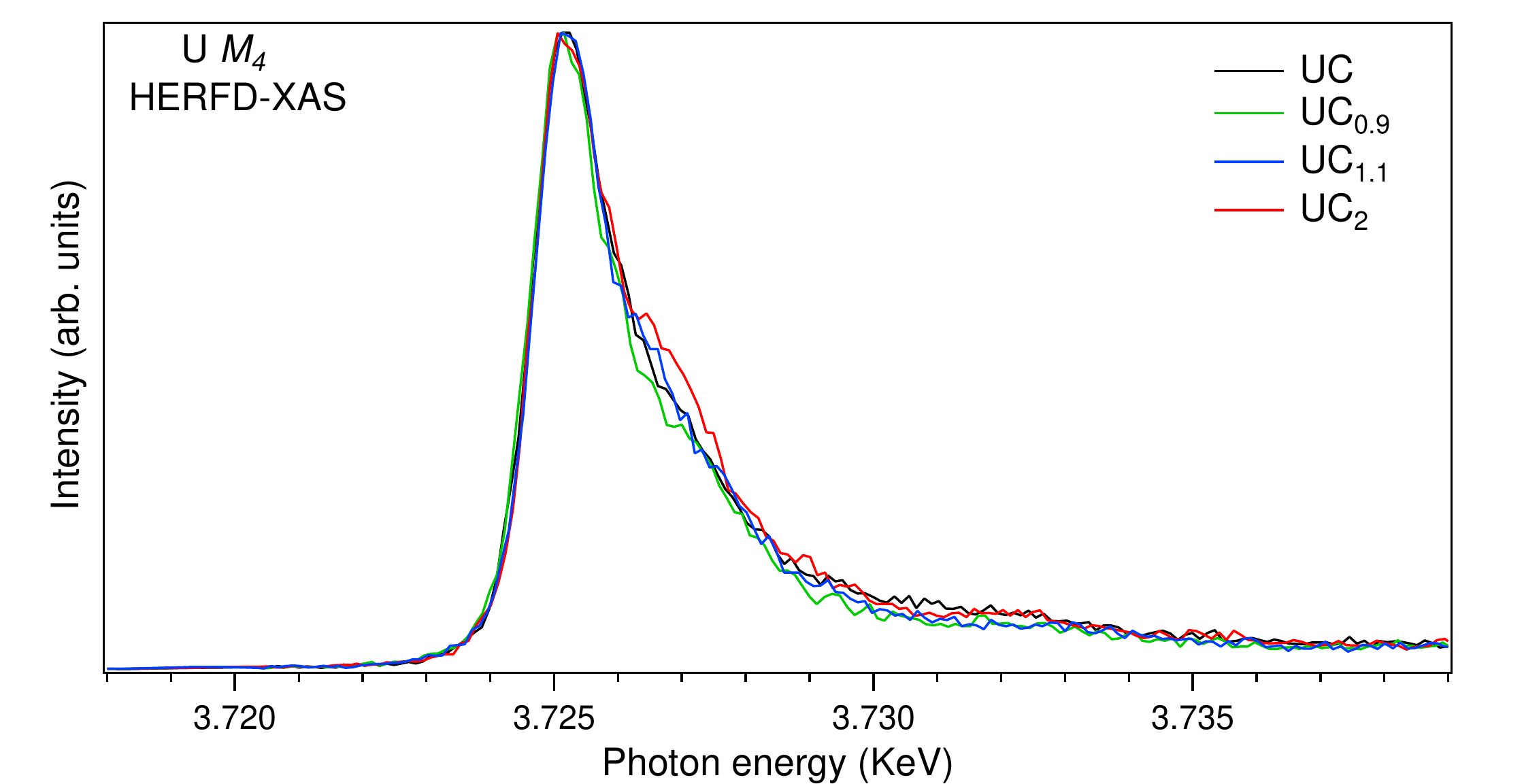}
\caption{U $M_4$ HERFD-XAS of UC$_{0.9}$, UC, UC$_{1.1}$ and UC$_{2}$.}
\label{UM4_HERFD}
\end{figure}

\begin{figure}
\includegraphics[width=\columnwidth]{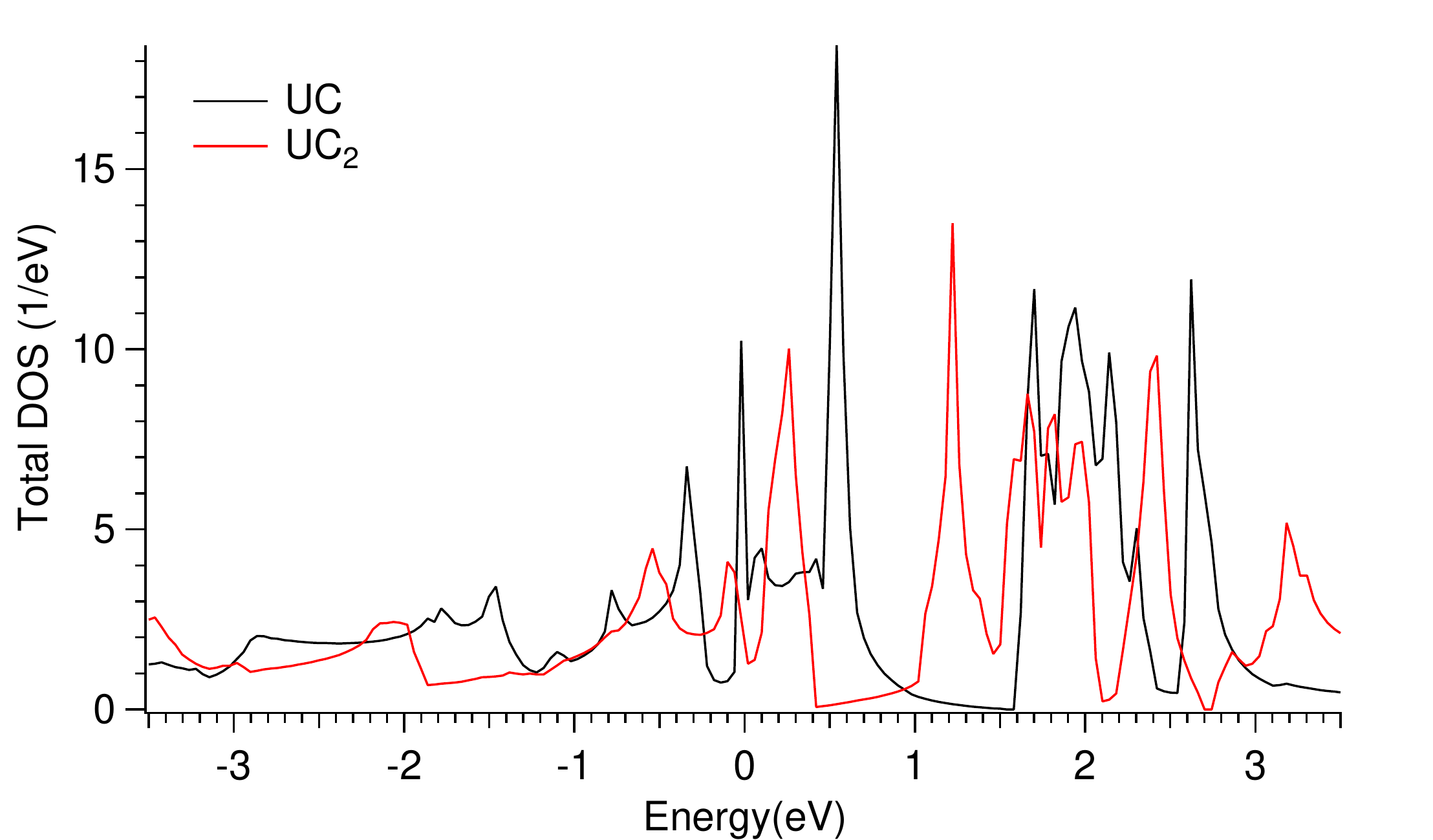}
\caption{Total DOS of UC and UC$_2$. Fermi level is at zero eV.}
\label{DOS_compare}
\end{figure}

\begin{figure}
\includegraphics[width=\columnwidth]{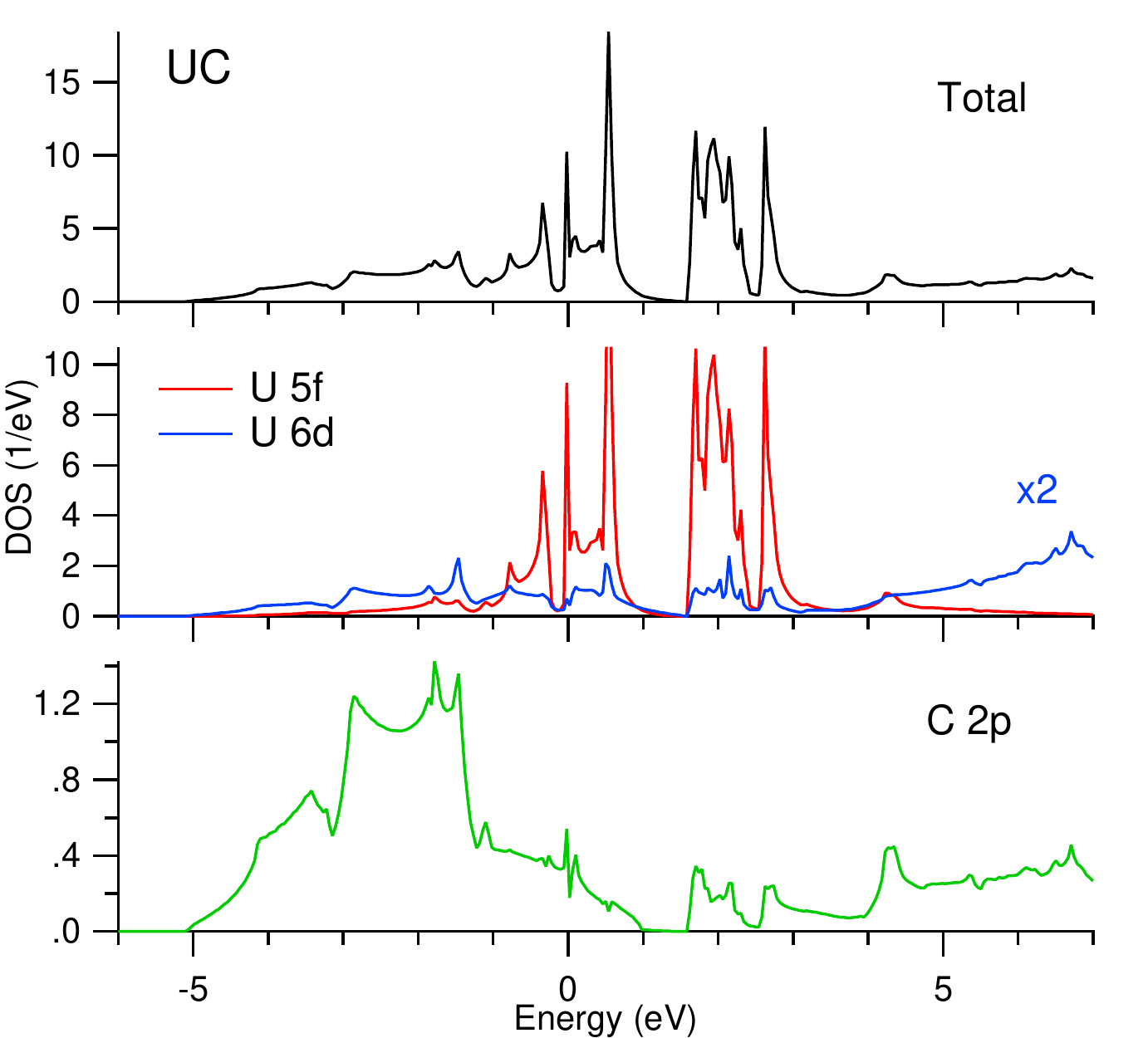}
\caption{Total and partial DOSs of UC. Fermi level is at zero eV.}
\label{DOS_UC}
\end{figure}

\begin{figure}
\includegraphics[width=\columnwidth]{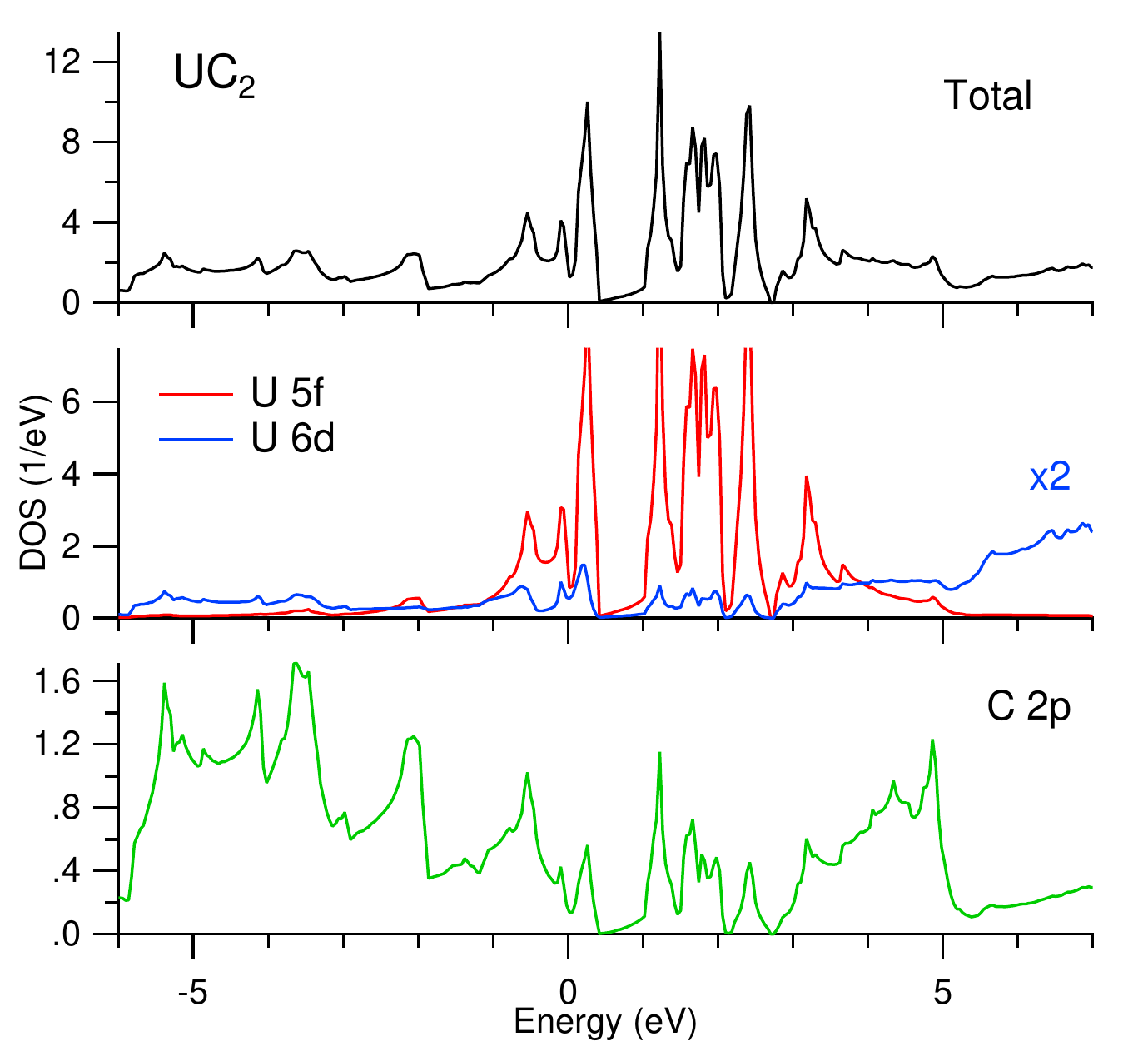}
\caption{Total and partial DOSs of UC$_2$. Fermi level is at zero eV.}
\label{DOS_UC2}
\end{figure}

On the other hand, the observed gradual relative-intensity increase of the shoulder in the U $M_4$ HERFD spectra of carbides can be related to the increase in the ligand/anion $2p\rightarrow$U $5f$ charge transfer and be a result of the multi-configurational contribution ($5f^{2}$ + $5f^{3}\underline{\upsilon}^{1}$ +  $5f^{4}\underline{\upsilon}^{2}$ + ..., where $\underline{\upsilon}$ stands for an electronic hole in the valence band) in the ground and final states of the spectroscopic process (as discussed in Ref. \cite{Butorin05}). That would mean a higher degree of such a charge-transfer for UC$_2$ compared to UC. Indeed, the full-relativistic DFT+$U$ calculations using the FPLO code (see Figs.~\ref{DOS_compare},\ref{DOS_UC},\ref{DOS_UC2}) support this conclusion based on their orbital population analysis. The U $5f$ occupancy was found to somewhat increase upon going from UC to UC$_2$.

The FPLO-calculated total DOS of UC and UC$_2$ is shown in Fig.~\ref{DOS_compare} and partial DOSs in Figs.~\ref{DOS_UC} and \ref{DOS_UC2}, respectively. Both for UC and UC$_2$, a clear energy separation into two groups of states above the Fermi level appears as a result of SOC, the observed splitting happens closer to the Fermi level in UC$_2$ than in UC. This again indicates the importance of taking into account SOC in the calculations of the electronic structure of actinide carbides. The prominent difference between UC and UC$_2$ is a wider spread of the occupied C $2p$ states in the valence band of UC$_2$ as a result of the hybridization with the U $6d$ states. At the same time, DOS at the Fermi level is significantly lower in UC$_2$ as compared to UC, thus affecting the thermodynamic properties and making UC more favorable material for the carbide nuclear fuel.

\section*{Conclusion}
The results of the present study indicate that the analysis of the effects of SOC and U is important for the proper description of the electronic structure of actinide carbides. The employment of x-ray spectroscopic methods with the enhanced energy resolution and sensitivity (such as HERFD-XAS at actinide $M_{4,5}$ edges) is necessary to spot small but important changes in the carbide properties with varying C content, such as the charge transfer and $5f$ occupancy, which have influence on the thermodynamic properties and eventually on the performance of the carbide nuclear fuel.

\section*{Methods}
\subsection*{Experimental}
The samples were prepared by arc melting of relevant proportions of metallic uranium (depleted uranium, Framatome, 99.9\%) and graphite pieces (Mersen) to reach the target stoichiometry of UC$_{0.9}$, UC, UC$_{1.1}$ and UC$_{2}$. To insure a good homogeneity, they were turned and re-melted four times. Each melting was performed under an argon pressure of about 0.8 bar, after three vacuum / Ar purges. All samples were characterized by x-ray diffraction (XRD), using a Bruker D8 Advance diffractometer with monochromatic Cu $K\alpha$ radiation. XRD patterns were refined by the Rietveld method, using the FullProf software \cite{Rodriguez}. The initial structural models for UC and UC$_{2}$ were taken from Ref. \cite{Austin}.

\begin{figure}
\includegraphics[width=\columnwidth]{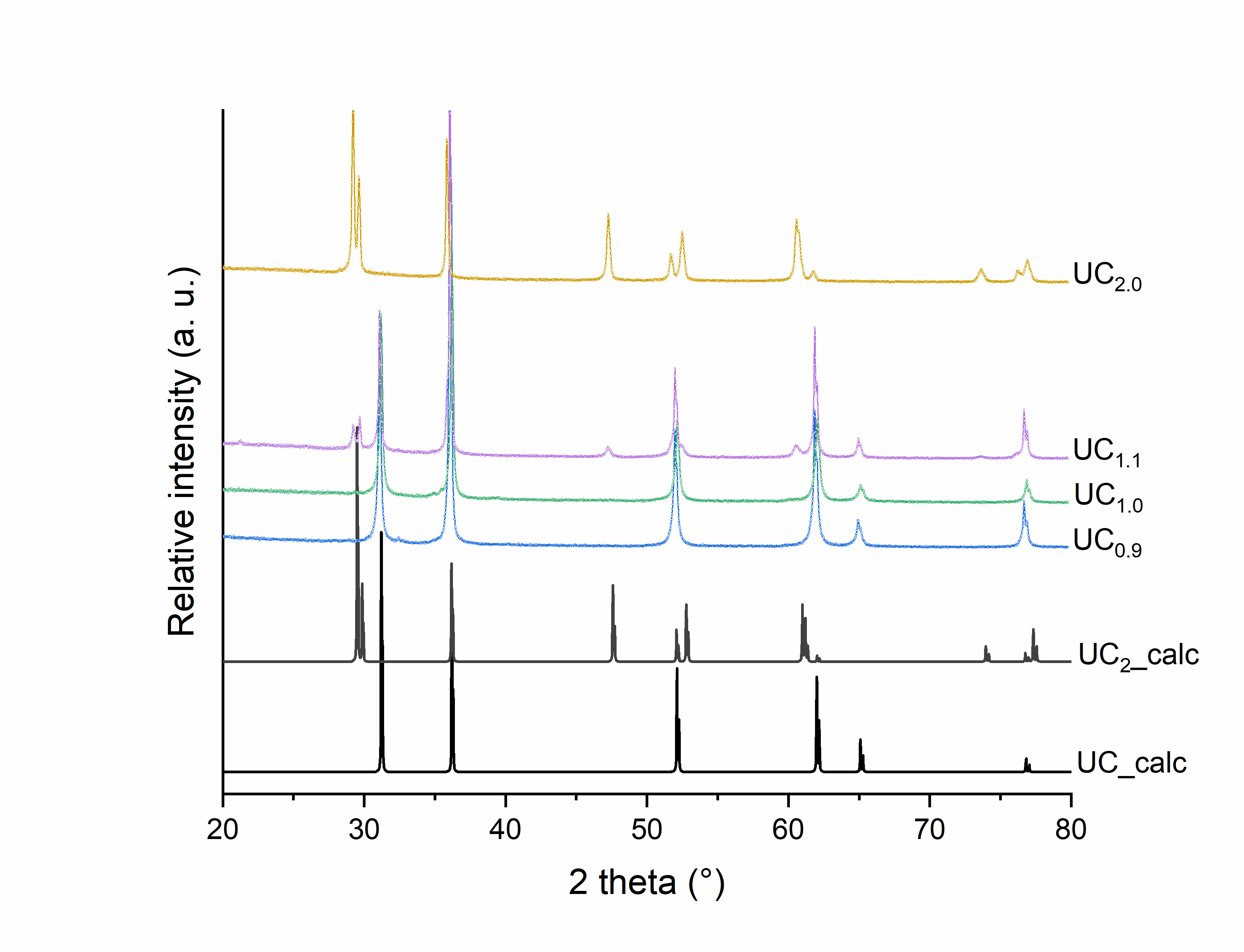}
\caption{Experimental x-ray diffraction patterns of the UC$_{0.9}$, UC, UC$_{1.1}$ and UC$_{2}$ samples compared to the calculated ones for Cu $K\alpha_{1,2}$ radiation based on the atomic positions of UC and UC$_{2}$ given in Ref. \cite{Austin}}
\label{XRD}
\end{figure}

A precise x-ray diffraction analysis (Fig.~\ref{XRD}) revealed that pure samples were obtained for the stoichiometric compositions UC$_{1.0}$ and UC$_{2.0}$ whereas mixtures of UC and $\alpha$-U and UC and UC$_{2}$ were obtained for UC$_{0.9}$ and UC$_{1.1}$ respectively. The refined lattice parameters were found consistent with those reported in the literature for stoichiometric uranium carbides, with $a$=4.956(1) {\AA} for cubic UC (NaCl-type) and $a$=3.524(1) {\AA} and $c$=5.991(1) {\AA} for tetragonal UC$_{2}$ (CaC$_{2}$-type).

The measurements in the energy range of the U $3d$ and $2p$ x-ray absorption edges were carried out at the CAT-ACT beamline \cite{Zimina} of the KARA (Karlsruhe research accelerator) facility in Karlsruhe, Germany. The incident energies were selected using the $<$111$>$ reflection from a double Si-crystal monochromator. The XAS scans were measured in the HERFD mode using an x-ray emission spectrometer \cite{Zimina}. Only one crystal-analyzer of the spectrometer was used in all the measurements. The sample, analyzer crystal and photon detector were arranged in the vertical Rowland geometry. The U HERFD spectra at the $M_{4}$ ($3d_{3/2}\rightarrow5f_{5/2},7p$ transitions) edge were obtained by recording the outgoing photons with an energy corresponding to the maximum of the  U $M\beta$ ($4f_{5/2}\rightarrow3d_{3/2}$ transitions) x-ray emission line, as a function of the incident energy, and the U HERFD at the $L_{3}$ ($2p_{3/2}\rightarrow6d,7s$ transitions) edge was recorded at the maximum of the U $L\alpha$ ($3d_{5/2}\rightarrow2p_{3/2}$ transitions) line. The right emission energy was selected using the spherically bent Si$<$220$>$ crystal-analyzer (with 1 m bending radius) aligned at 75$^\circ$ Bragg angle for the measurements at the U $M_{4}$ edge and Ge$<$777$>$ at 77$^\circ$ Bragg angle for the measurements at the U $L_{3}$ edge. The HERFD data were recorded at one emission energy and the spectrometer was not moved between the scans. The spectral intensity was normalized to the incident flux. The total energy resolution was estimated to be $\sim$0.7 eV at the U $M_{4}$ edge and $\sim$2.6 eV at the U $L_{3}$ edge.

Samples were prepared and sealed in a special argon-filled container at the licensed laboratory of HZDR and were transported to KARA under inert conditions. All samples were mounted in the form of pellet pieces within triple holders with 8 $\mu$m Kapton window on the front side, serving as first confinement. Three of such holders were mounted in one larger cell, with 13 $\mu$m Kapton window on the front side. The second confinement chamber was constantly flushed with He. The entire spectrometer environment was contained within a He box to improve signal statistics.

The measurements in the energy range of the C $K$ ($1s\rightarrow2p$ transitions) edges of UC were performed at beamline 5.3.1 of the MAXlab \cite{Denecke}. C $K$ XAS data were measured in the total fluorescence yield (TFY) mode using a multichannel-plate detector. The incidence angle of the incoming photons was close to 90$^\circ$ to the surface of the sample. The monochromator resolution was set to $\sim$350 meV during the measurements.

\subsection*{Computational details}
To apply the method of density functional theory taking into account the $5f-5f$ Coulomb interaction $U$ (DFT+$U$), the full-potential local orbital (FPLO ver. 21.00-61) code (Ref.\cite{Koepernik}; [www.FPLO.de]) was used. The calculations were performed in the full-relativistic mode. The exchange correlation potential was in the form of Perdew, Burke, and Ernzerhof (PBE) \cite{Perdew}. The band structure was calculated in the generalized gradient approximation (GGA). The default basis definitions for uranium and carbon atoms were applied, where the core electrons for uranium are up to the $5p$ level while $5d6s6p7s7p6d5f$ electrons are treated as valence ones and the levels up to $8s$ are also included in the basis. For carbon, the $1s$ electrons are treated as semi-core and the $n$=3 levels are included in the basis in addition to valence $2s$ and $2p$. The Coulomb interaction $U$ and Hund's coupling $J$ parameters were set to 2.5 eV and 0.5 eV, respectively, for the $5f$ shell. There is rather a consensus among researchers on these values for UC. The calculations were performed for the non spin-polarized case. The 40x40x40 $k$-point mesh was used for UC and the 30x30x30 one for UC$_2$. As convergence conditions, 10$^{-10}$ for density and 10$^{-8}$ Ry for the total energy were applied. The calculations were performed for experimental structures of UC and UC$_2$ \cite{Austin} without the relaxation procedure.

The experimental U $L_{3}$ HERFD-XAS and C $K$ XAS spectra of UC are also compared with the results of the XAS calculations using OCEAN which is the \textit{ab-initio} DFT (PBE-GGA in this case) + Bethe-Salpeter equation (BSE) code for the calculations of core-level spectra.\cite{Vinson01,Vinson02} The code allows one to take into account the interaction of the valence-band electrons with the U $2p$ (C $1s$) core-hole and screening effects in the calculations of the U $L_3$ (C $K$) XAS spectra, respectively. The DFT+$U$ approach was used with the help of Quantum Espresso v6.3 (Ref.\cite{Giannozzi01,Giannozzi02}; [www.quantum-espresso.org]) and the $U$ and $J$ values were set to 2.5 eV and 0.5 eV, respectively, for the uranium $5f$ electrons. The norm-conserving PBE pseudopotential for carbon was taken from the PseudoDojo database \cite{vanSetten}. The stringent version of the C pseudopotential of the valence $2s^22p^2$ configuration was used. The norm-conserving PBE pseudopotential for uranium was generated with the ONCVPSP v4.0.1 package \cite{Hamann} for the valence $6s^26p^67s^26d^15f^3$  configuration using the PseudoDojo approach. The plane-wave cut-off energy was set to 65 Ry. The convergence threshold for density was 1.1x10$^{-10}$ Ry. The $k$-point grid for the calculation of the ground and final states as well as the real space mesh were 10x10x10 for UC and 10x10x6 for UC$_2$. The 2x2x2 $k$-point grid was used for the screening part of the calculations. The setting for the DFT and screening energy ranges were chosen to be 90 and 100 eV, respectively.

%/////////////////////////////////////////////////////////
%------------------- Acknowledgements ---------------------
%/////////////////////////////////////////////////////////
\section*{Acknowledgments}
S.M.B. acknowledges support from the Swedish Research Council (research grant 2017-06465). S.B., L.A. and K.O.K. acknowledge support by the European Research Council under grant No. 759696. The computations and data handling were enabled by resources provided by the Swedish National Infrastructure for Computing (SNIC) at National Supercomputer Centre at Link\"{o}ping University partially funded by the Swedish Research Council through grant agreement no. 2018-05973.

\section*{Author contributions statement}
S.M.B. conceived, planned and conducted the experiments, performed the calculations and analyzed the results, S.B. and L.A. conducted experiments and analyzed the results, A.B. A.R. and T.V. conducted the experiments, K.O.K. planned and conducted the experiments, S.W. and O.T. prepared and characterized the samples. All authors reviewed the manuscript.

\section*{Additional information}
\subsubsection*{Competing interests}
The authors declare no competing interests.

\subsubsection*{Data availability}
The datasets generated during and/or analysed during the current study are available from the corresponding author on reasonable request.

%\begin{tocentry}
%\includegraphics{TOC_SBM.pdf}
%A theoretical overview of the core-to-core resonant inelastic x-ray scattering of actinide dioxides is provided using crystal-field multiplet theory. The calculations allowed for a general %analysis of high-energy-resolution fluorescence-detected x-ray absorption (HERFD-XAS) spectra and their comparison with conventional XAS.
%\end{tocentry}

%/////////////////////////////////////////////////////////
%--------------------- Bibliography ----------------------
%/////////////////////////////////////////////////////////

\bibliography{U_carbides_II}

\end{document}